\documentstyle[floats,aps,epsf,prl]{revtex}

\begin{document}
\draft
\twocolumn[\hsize\textwidth\columnwidth\hsize\csname @twocolumnfalse\endcsname

\title{Quantum transport through ballistic cavities: soft vs. hard
  quantum chaos}

\author{Bodo Huckestein$^{1}$\cite{p_address}, Roland Ketzmerick$^2$, and Caio
  H. Lewenkopf$^3$}

\address{$^1$ Institut f\"ur Theoretische Physik, Universit\"at zu
  K\"oln, D-50937 K\"oln, Germany\\
  $^2$ Max-Planck-Institut f\"ur Str\"omungsforschung and Institut
  f\"ur Nichtlineare Dynamik der Universit\"at G\"ottingen,\\
  Bunsenstra{\ss}e 10, D-37073 G\"ottingen, Germany\\
  $^3$ Instituto de F\'{\i}sica,
  Universidade do Estado do Rio de Janeiro, \\
  R. S\~ao Francisco Xavier, 524, CEP 20559-900 Rio de Janeiro,
  Brazil}

\date{6 August 1999}

\maketitle

\begin{abstract}
  We study transport through a two-dimensional billiard attached to
  two infinite leads by numerically calculating the Landauer
  conductance and the Wigner time delay. In the generic case of a
  mixed phase space we find a power law distribution of resonance
  widths and a power law dependence of conductance increments
  apparently reflecting the classical dwell time exponent, in striking
  difference to the case of a fully chaotic phase space.
  Surprisingly, these power laws appear on energy scales {\em below}
  the mean level spacing, in contrast to semiclassical expectations.
\end{abstract}
\pacs{PACS numbers: 05.45.+b,05.60.Gg,72.20.Dp,73.23.Ad}
\vskip2pc

]

Advances in the fabrication of semiconductor heterostructures and
metal films have made it possible to produce two dimensional
nanostructures with a very low amount of disorder
\cite{Kouwenhoven98}. At low temperatures,
scattering of the electrons happens mostly at edges of the structures
with the electrons moving ballistically between collisions with the
boundary. Theoretical and experimental investigations have shown
that the spectral and transport properties of such quantum coherent
cavities, commonly called ``billiards'', depend strongly on
the nature of their classical dynamics. In particular, integrable and
chaotic systems were found to behave quite differently
\cite{marcus93Chang94,BJS93}.

\begin{figure}[b]
  \begin{center}
    \epsfxsize=6.2cm
    \leavevmode
    \epsffile{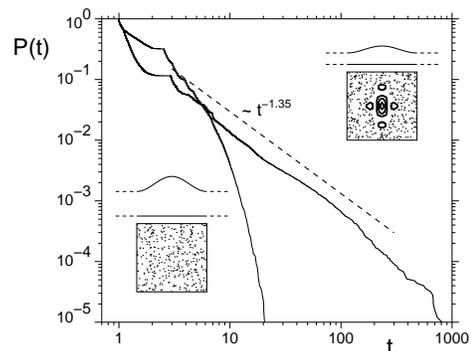}
        \caption{Classical dwell time probability $P(t)$, for the cases
          of mixed (thick line) and fully chaotic (thin line) dynamics
          with $t$ in units of the traversal time. The insets show the
          shape of the billiard with attached leads and a Poincar\'e
          surface of section of the closed billiard for the mixed
          (upper right) and chaotic (lower left) case.}
    \label{fig:class}
  \end{center}
\end{figure}

Generic billiards are neither integrable nor ergodic \cite{MM74}, but
have a mixed phase space with regions of regular
as well as chaotic dynamics \cite{LL92}. Their dynamics is much richer
than in either of the extreme cases, as phase space has a hierarchical
structure at the boundary of regular and chaotic motion.  In
particular, this leads to a trapping of chaotic trajectories close to
regular regions with a probability $P(t) \sim t^{-\beta}$ for $t>t_0$,
to be trapped longer than a time $t$,
with $t_0$ of the order of a few traversal times \cite{powerlaw}.
The exponent $\beta > 1$ depends on system and parameters
with typically $\beta \approx 1.5$ \cite{powerlaw}. This power-law
trapping in mixed systems is in contrast to the typical exponentially
decaying staying probability of fully chaotic systems (see
Fig.~\ref{fig:class}).

Recently, it was shown semiclassically employing the diagonal
approximation that the variance of conductance increments 
(for a small dc bias voltage) over small
energy intervals $\Delta E$ grows as \cite{Ket96,Lai92}
\begin{equation}
\label{variance}
\Delta g^2(\Delta E) \equiv \langle [g(E+\Delta E) - g(E)]^2 \rangle_{E}
\sim |\Delta E|^{\beta} ,
\end{equation}
for mixed systems if $\beta < 2$.  This is in strong contrast to an
increase as $(\Delta E)^2$ in the case of fully chaotic systems
\cite{BJS93}. The semiclassical approximation requires $\Delta E$ to
be {\em larger} than the mean level spacing $\Delta$, corresponding to
the picture that quantum mechanics can follow the classical power law
trapping at most until the Heisenberg time $t_H=h/\Delta$ \cite{qpr}.
In the semiclassical approximation the graph of $g$ vs.\ $E$ has the
statistical properties of fractional Brownian motion with a fractal
dimension $D=2-\beta/2$ \cite{Ket96}.  Fractal conductance
fluctuations have indeed been found in experiments on gold wires
\cite{Hea96} and semiconductor nanostructures \cite{Sea98b} and 
numerically for the quantum separatrix map \cite{sep}.

In this Letter, we numerically study quantum transport through a 
simple cavity, the cosine billiard \cite {cosinus}
(see insets of Fig.~\ref{fig:class}).
Although we observe completely different behavior for the mixed and
fully chaotic cases in the semiclassical regime of many (45)
transmitting modes, we find in the mixed case no indication of
fractality or fractional Brownian motion behavior of the graph
$g$ vs.\ $E$. This is the first surprise, as it is
in contrast to the above mentioned semiclassical
\cite{Ket96}, experimental \cite{Hea96,Sea98b}, and numerical
\cite{sep} works.  Instead, the conductance is characterized by
narrow isolated resonances, with the classical exponent $\beta$
appearing in a power law distribution of resonance widths
{\em smaller} than the mean level spacing. This leads to a scaling
of $\Delta g^2(\Delta E)$ in agreement with the semiclassically derived
Eq.~(\ref{variance}), however, only on scales {\em below} the mean
level spacing.
This surprising result contradicts the semiclassical intuition that 
quantum mechanics may mimic classical properties at most until the
Heisenberg time corresponding to energy scales {\em above} the mean level 
spacing.
At present, there is no explanation for these numerical results.  They
show that even with a detailed (semiclassical) knowledge of the universal
chaotic regime as well as the integrable case at hand
we are just at the beginning of understanding the
quantum properties of {\em generic} Hamiltonian systems.

The cosine billiard \cite {cosinus} is defined by two hard walls
at $y=0$ and $y(x) = W + (M/2)( 1 - \cos(2\pi x/L))$, 
for $0\leq x\leq L$, with two semi-infinite perfect leads of width
$W$ attached to the openings of the billiard at $x=0$ and $x=L$
(see insets of Fig.~\ref{fig:class}).
By changing the parameter ratios $W/L$ and $M/L$ the stability of
periodic orbits associated with the billiard
can be changed, allowing a transition from a mixed to a predominantly
chaotic phase space. Note, that in the mixed case the leads couple to
the chaotic part of  phase space only.

The $S$-matrix of the system has been calculated by the recursive
Green's function method after expanding the two-dimensional wave
function in terms of local transverse energy
eigenfunctions \cite{MZH94}.  In the numerical calculations, it was
checked that a sufficient number of modes in the expansion in
transverse eigenmodes was kept and that the lattice constant in
$x$-direction was sufficiently small.  For a given energy $E_F=\hbar^2
k_F^2/2m$, $N$ modes in the leads are transmitting, with $k_F
W/\pi\geq N$.  We turn from the $S$-matrix to the
experimentally relevant conductance at small dc bias voltage
using the Landauer formula, $G=e^2/h\mbox{Tr}\,
(tt^\dagger)$, where $t$ is the transmission matrix.  Spectral
information is contained in the Wigner-Smith time delay $\tau=-i\hbar
\mbox{Tr}\,(S^\dagger dS/dE)/2N$, where $2N$ is the dimension of the
$S$-matrix.  All energies in this paper are given in units of
$\hbar^2\pi^2/(2mW^2)$.

\begin{figure}
  \begin{center}
    \epsfxsize=7.4cm
    \leavevmode
    \epsffile{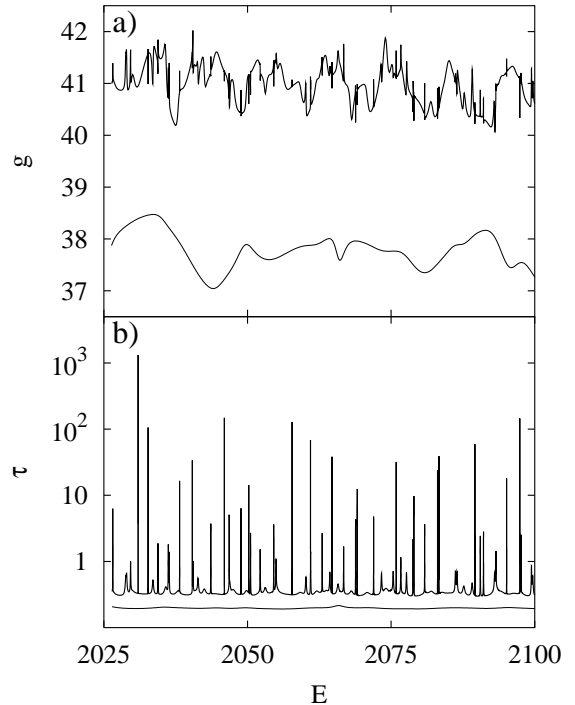}
    \caption{Dimensionless conductance $g$ (a) and Wigner time delay
      $\tau$ (b) for $N=45$ propagating modes. Thick (thin) lines for
      mixed (fully chaotic) case.}
    \label{fig:cond}
  \end{center}
\end{figure}

Figure~\ref{fig:cond} shows the dimensionless conductance $g=G(h/e^2)$
and the Wigner-Smith time delay $\tau$ [in units of
$2mW^2/(\hbar\pi^2)$] for parameters corresponding to a mixed phase
space ($W/L=0.18$, $M/L=0.11$) and a chaotic phase space with no
apparent stability island ($W/L=0.36$, $M/L=0.22$) for $N=45$
transmitting modes.  The differences are quite dramatic.  For the
fully chaotic case, both quantities are smooth functions of energy and
in good agreement with semiclassical theory (see below). While
the average values are comparable, many sharp
isolated resonances on top of a smooth background are visible in the
mixed case \cite{Guarneri}, also in contrast to the
semiclassically predicted fractional Brownian motion.  The simple
explanation that these narrow resonances are related to quantum
tunneling into the islands of regular motion \cite{S93} does not apply
here, as the phase space volume of stable islands is about 5\%, while
the narrow resonances (below the mean level spacing) make up about 18\% of
all states associated with the billiard. This roughly corresponds to
the phase space volume around the stable islands where trapping of
chaotic trajectories occurs.

\begin{figure}
  \begin{center}
    \epsfxsize=7.6cm
    \leavevmode
    \epsffile{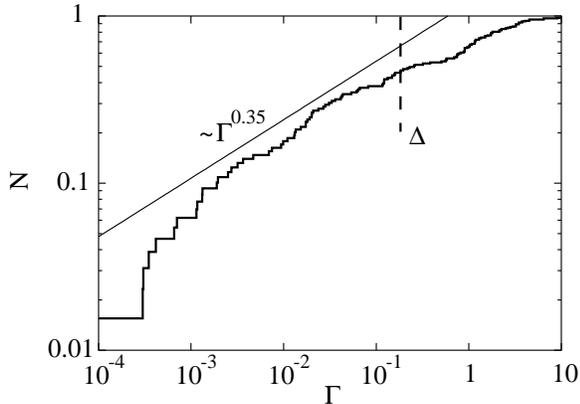}
    \caption{Normalized cumulative distribution of resonances widths
      $N(\Gamma)$ for the mixed case. The thin line serves as a
      guide to the eye. The mean level spacing $\Delta$ is shown
      in the figure.}
    \label{fig:dist}
  \end{center}
\end{figure}

In order to analyze the narrow resonances in the mixed case, it is
convenient to examine the Wigner-Smith time delay.
Each resonance in the time delay has the Breit-Wigner shape,
characterized by a width $\Gamma_i$ and a height $\tau_i$ situated at
an energy $E_i$ on top of a smooth background. We find our data well
described by
\begin{equation}
  \label{eq:bw}
  \tau(E) = \sum_i \tau_i \,\frac{\Gamma_i^2/4}{(E-E_i)^2+\Gamma_i^2/4}
             + \tau_{\mbox{\scriptsize smooth}}(E),
\end{equation}
with $\tau_{\mbox{\scriptsize smooth}}(E)\propto E^{-1/2}$.  Since the
phase shift through a resonance is $2\pi$, width and height are
related by $\tau_i\Gamma_i=2/N$. The energy was initially sampled on
an equidistant grid and subsequently refined in order to resolve the
sharp resonances.  Only resonances with a $\Gamma\lesssim 10^{-3}$,
i.e.\ much smaller than the initial grid are lost.  As a result we can
numerically construct the cumulative distribution $N(\Gamma)$ of
resonance widths, corresponding to the probability of finding a
resonance smaller than $\Gamma$ (Fig.~\ref{fig:dist}). The
distribution is very broad, spanning 5 orders 
of magnitude, and is approximately a power law $N(\Gamma) \approx a
\Gamma^r$, with $r \approx 0.35$, over a wide range {\em below} the
mean level spacing $\Delta = 0.176$.

The consequences of this broad distribution of resonance widths for
the variances of conductance and time delay increments are studied
now. For $\Delta E \ll \Delta$ correlations between different isolated
resonances ($\Gamma \ll \Delta$) do not contribute to the variance
$\Delta g^2$, which then is governed by the distribution of resonance
widths. Each resonance is reflected in the conductance,
\begin{equation}
  \label{eq:g_mod}
  g(E) = g_{\mbox{\scriptsize smooth}}(E) + \sum_{i=1}^{N_R} \delta
  g_i(E),
\end{equation}
where $\delta g_i(E)$ is a function of the width $\Gamma_i$ and
the typical height
$\widetilde{g}_i$. $N_R$ is the number of resonances with
$\Gamma_i < \Delta$ in the energy interval $E_B-E_A$ over which we
take the average.
The variance of the increments of a single resonance is given by
\begin{eqnarray}
  \label{eq:single_est}
  \lefteqn{\langle [\delta g_i(E+\Delta E) -\delta g_i(E)]^2\rangle_E
    =}\nonumber \\
  & & \frac{\widetilde{g}_i^2\Gamma^{\vphantom{2}}_i}{E_B-E_A} \left\{
    \begin{array}{ccc}
      b_i(\Delta E/\Gamma_i)^2 &,& \Delta E \ll \Gamma_i \\
      1 &,& \Delta E \gg \Gamma_i
    \end{array}\right.,
\end{eqnarray}
which defines $\widetilde{g}_i$ and where $b_i$ is a numerical factor
of order unity. Since the distribution of widths is very broad, the strong
inequalities are almost always fulfilled in the sum over resonances.
Splitting this sum into resonances smaller and larger than $\Delta E$,
we get
\begin{equation}
  \label{eq:var_sum}
  \Delta g^2(\Delta E) \approx \frac{1}{E_B-E_A}
  \left(\sum_{\Gamma_i<\Delta E}\!\!\! \widetilde{g}_i^2\Gamma^{\vphantom{2}}_i +
    \sum_{\Gamma_i>\Delta E}\!\!\!
    b^{\vphantom{2}}_i \widetilde{g}_i^2 \frac{(\Delta E)^2}{\Gamma_i}\right).
\end{equation}
Replacing the sums by integrals over the density of widths
$n(\Gamma)\approx ar\Gamma^{r-1}$ and neglecting the weak fluctuations
of $g_i$ and $b_i$ as compared to $\Gamma_i$, we can estimate for
small $\Delta E$,
\begin{equation}
  \label{eq:dg_est}
    \Delta g^2(\Delta E) \propto \langle \widetilde{g}^2\rangle^{\vphantom{2}}
    \frac{N_R}{E_B-E_A} a |\Delta E|^{1+r}.
\end{equation}
where $\langle \cdots \rangle$ stands for the average over isolated
resonances.
A power law distribution of resonances thus leads to a power law
increase of the variance of conductance increments with the exponent
given by $1+r$.

\begin{figure}
  \begin{center}
    \epsfxsize=7.6cm
    \leavevmode
    \epsffile{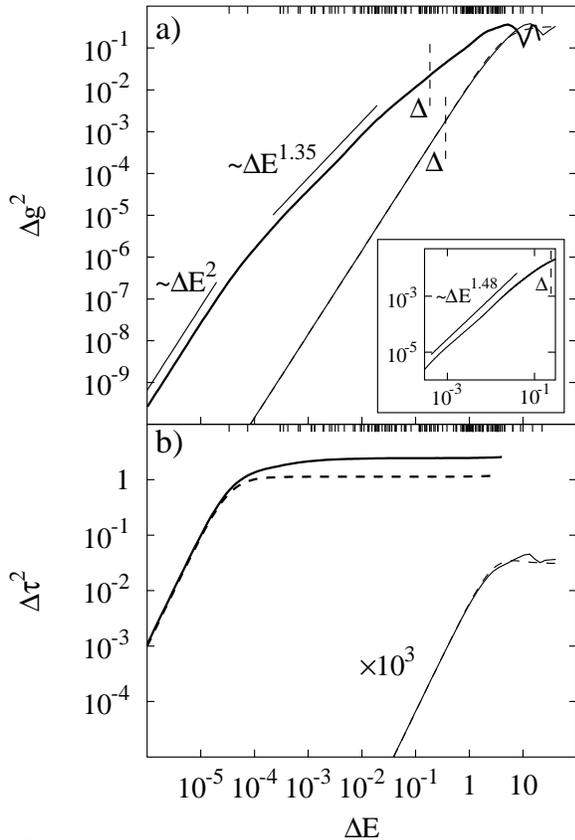}
    \caption[Variance]{Variance of the increments of the conductance $\Delta
      g^2(\Delta E)$ (a) and Wigner time delay $\Delta \tau^2(\Delta
      E)$ (b). The thick (thin) lines correspond to the mixed (fully
      chaotic) case. The thin dashed lines are fits to the
      semiclassical expressions for the chaotic case. The thick dashed
      line in Fig.~b) is the contribution of the sharpest resonance to
      the variance for the mixed case. For comparison, the ticks on
      the upper borders mark the widths of the individual resonances.
      The inset in a) shows the corresponding data for a quantum graph
      modeling a mixed phase space ($\beta=1.48$) and was
      provided by the authors of ref.~\cite{KW99}. The mean level
      spacing $\Delta$ is shown in the figures.}
    \label{fig:variance}
  \end{center}
\end{figure}

Fig.~\ref{fig:variance}~a) shows the variances of the
conductance increments.
On scales smaller than the minimum resonance width the
variance increases quadratically, as expected.
On larger scales we find the power law Eq.~(\ref{eq:dg_est}).
This result coincides with the semiclassically derived
Eq.~(\ref{variance}) with $r=\beta-1$,
however, only on scales {\em below} the mean level spacing.
At present, there is no explanation why the classical exponent $\beta$
appears on such small energy scales.
Remarkably, on scales above the mean level spacing the correlation
energy for the conductance fluctuations is given by the Weisskopf width
$\Gamma_W\approx 2$, as in the fully chaotic case (see below).

The variance of increments of the time delay are shown in
Fig.~\ref{fig:variance}~b).  Since the variance $\Delta\tau^2(\Delta
E)$ measures the square of the resonance peak height in the time
delay, in the mixed case, they are completely dominated by the
sharpest resonance, once the energy exceeds the minimum resonance
width. Thus, in contrast to fully chaotic systems, in mixed systems
the scale of the correlations of the time delay is the smallest
resonance width and not the Weisskopf width $\Gamma_W$.

For comparison in Fig.~\ref{fig:variance} we also show the results for
$\Delta g^2(\Delta E)$ and $\Delta\tau^2(\Delta E)$ for the fully
chaotic case. They are characterized by single scales
$\Gamma_{g}\approx4.8$ and $\Gamma_{\tau}\approx 3.8$ and are in good
agreement with semiclassical results \cite{BJS93,Eck93}. The chaotic
case can also be described by the random matrix theory (RMT)
\cite{GMW98}, whose results coincide with the cited semiclassical ones
for $N\gg 1$ \cite{Leh95}. In the absence of direct processes, random
matrix theory predicts a single correlation scale, known as Weisskopf
correlation width, $\Gamma_W=\Delta/2\pi\sum_c T_c$, where the sum
runs over all channels $c$ with transmission probability $T_c$
\cite{Lew91}. Approximating $\sum_c T_c$ by twice the average
dimensionless conductance we obtain $\Gamma_W \approx 4.2$, in
agreement with the numerical values within the statistical accuracy.
Before concluding, it is worthwhile to stress that depending on $N$
and the coupling to the leads, quantum chaotic scattering can also
exhibit isolated resonances. Their width distribution, however,
follows a $\chi^2$-distribution with $N$ degrees of freedom
\cite{GMW98}, rather than power law.

In conclusion, we have shown that generic Hamiltonian systems, which
have regular as well as chaotic phase space regions, differ
drastically in the Landauer conductance and Wigner time delay from
fully chaotic systems.  We find many isolated narrow resonances with a
power law distribution of their widths accompanied by a power law
increase of the variance of conductance increments.  Both power laws
appear to be connected to the classical power law trapping,
surprisingly they only appear on scales {\em below} the mean level
spacing. 
Similar unexplained power laws are found in 
recent studies using quantum graphs \cite{KS97} modeling a mixed phase
space (see inset of Fig.~\ref{fig:variance}~a) \cite{KW99}.
Further research on the quantum signatures of classically
mixed systems is urgently needed.

R.K. thanks I. Guarneri for many stimulating discussions.  We thank L.
Hufnagel, F. Steinbach, and M. Weiss for the inset of
Fig.~\ref{fig:variance}~a). This work was supported by the DFG and
the ITP at UCSB (B.H.), CNPq and PRONEX (C.H.L.), and the ICTP in
Trieste (B.H and C.H.L.). 

{\em Note added.\/}---The authors of Ref.~\cite{KW99} have informed us
that there are quantum graphs where the classical and quantum
exponents do not agree.

\end{document}